\documentclass[preprintnumbers,amsmath,amssymb,floatfix,11pt,prd,onecolumn,
superscriptaddress,nofootinbib]{revtex4}
\usepackage{graphicx}
\usepackage{epsfig}
\usepackage{bm}
\usepackage{amsfonts}
\begin{document}

\title{Warm Intermediate Inflation in  $F(T)$ Gravity}

\author{\textbf{Mubasher Jamil}}\affiliation{Department of Mathematics,
School of Natural Sciences (SNS),National University of Sciences and
Technology (NUST), H-12, Islamabad, Pakistan}
\author{\textbf{Davood Momeni}}\affiliation{Eurasian International Center for Theoretical Physics, Eurasian National University, Astana 010008, Kazakhstan}
\author{\textbf{Ratbay Myrzakulov}}\affiliation{Eurasian International Center for Theoretical Physics, Eurasian National University, Astana 010008, Kazakhstan}

\begin{abstract}
\textbf{Abstract:} We investigate warm intermediate scenario of the
cosmological inflation in $F(T)$ gravity in the limit of high
dissipation. The inflationary expansion is driven by the scalar
inflaton while the gravitational dynamics follow from the $F(T)$
gravity. We calculate the relevant inflationary observables such as
scalar-tensor ratio, power-spectrum indices of density perturbations
and gravitational waves and the e-folding parameter. We obtain a
ratio of slow-roll parameters to be constant. Our calculations
support the warm-intermediate inflationary scenario in a spacetime
with torsion. Moreover our results are compatible with the
astrophysical observations of cosmic microwave background and Planck
data.


\end{abstract}
\maketitle
\newpage
\section{Introduction}

The hypothesis of cosmological inflation in the early Universe is
useful in answering several fundamental questions of cosmology such
as why the energy-matter distribution in the Universe is
homogeneous; problem of fine-tuning of the initial velocities of the
energy-matter (the flatness problem), the magnetic monopole problem
and the horizon problem. Moreover the inflaton that drives the
inflation serves as the harbinger of seed fluctuations for later
large scale structure formation \cite{guth}. Observations of the
cosmic microwave background (CMB) and the large-scale structure
(LSS) are used to determine the spectrum of primordial seed
fluctuations. This makes CMB and LSS experiments the only probes of
the very early Universe. Current observations are in excellent
agreement with the basic inflationary predictions: The Universe has
an almost scale-invariant Gaussian power spectrum \cite{cole}. To
have a better view on  the essence of the late time cosmic
acceleration, one geometrical approach is to use the $f(R)$ gravity.
This theory was proposed by Buchdahl as a mathematical extension of
the Einstein gravity in a very simple and systematic form
\cite{buchdahl}. Other possibilities include a non-zero cosmological
constant or dynamical dark energy models \cite{lorenzo}. Different
aspects of dark energy models have been discussed recently (for a
review see \cite{review}).

Curvature is not a unique description of the gravity. Going beyond
the Riemannian geometry, one can use torsion as an alternative for
geometrical description of the gravity \cite{hayashi}. General
relativity is constructed using a symmetric Levi Civita connection
whereas teleparallel gravity is constructed from a skew-symmetric
Weitzenbock connection. In recent years attentions have been focused
on the generalizations of this idea as teleparallel gravity
\cite{ft1,ft11}. Different cosmological aspects of $F(T)$ gravity
such as resolution of dark energy and dark matter problems with
torsion have been discussed in literature (see \cite{ft2, paper} and
references therein). Also some black holes solutions are also
derived in $F(T)$ gravity \cite{ft}.

Conventional inflation model has two distinct stages of evolution.
The first stage is governed by the rapid accelerated expansion
driven by the inflaton with negligible kinetic energy and a stable
potential energy. In the second stage (called reheating), the
inflaton decays into matter and radiation fields which is a kind of
a hot Big Bang. The main problem is how we can join the universe
towards the end of this era, successfully. Although both stages are
driven by different physical mechanisms, the idea of 'warm
inflation' amazingly unifies them \cite{Barrow}. Here inflation is
described as a decay of the field into thermal component via weak
and strong dissipation regimes. In this scenario, it is assumed that
the radiation (produced during the inflation) keeps a constant
density. This constant energy density preserves the form of the
cosmological solution as a transition phase in the form of the
de-Sitter universe. To validate this proposal, we need a dissipative
formalism. The dissipation coefficient $\Gamma$ is necessary to
explain this heating phase completely. The full consistent  model to
construct this dissipation functions is fulfilled by quantum field
theory tools using a two stage mechanism, applied on the
interactions \cite{qft}.  We need warm inflationary epoch to stop
the inflation in a finite time. Dynamics of the warm inflation has
some interesting features. The main aspect is, during the radiation
production, some microscopic (micro-statistical) procedures happen.
The average time of this cascade process goes faster than the
velocity (here Hubble parameter) of the background. If we denote by
$\Gamma_a$ a typical decay rate of one of these process, than it
means that $\Gamma_a>H$. This implies that the quantum photon
productions slows-down the dynamics of the cosmological background.
It is a kind of the slow-roll approximation. This idea has
motivations from the tachyonic field as well \cite{setare}, the
Hawking radiation \cite{singleton} and via holographic principle
\cite{jamil}. The scale factor of the cosmological background
through the inflationary era must be a power law or exponential of
an intermediate form. Such forms of the scale factor possess exact
solutions for fields (vector, scalar,..). One important form
proposed is the following \cite{barrow1}
\begin{equation}
a(t)=\exp(A_1t^f),\quad 0<f<1, \quad A_1>0
\end{equation}
where $f$ and $A_1$ are constants. In this model, the expansion rate
is between de Sitter inflationary expansion $(a(t) = \exp(Ht))$, and
power-law inflationary expansion $(a(t) = t^\alpha, \alpha
> 1)$. The intermediate inflationary scenario has been proposed  as
  solution for a constrained inflaton potential function in the
form $V(\phi)\propto\phi^{-m},\ \ m=4(f^{-1}-1),0<f<1$
\cite{campo1}. This form of the scale factor is also motivated from
string/M theory as the cosmological solutions of the weak field of
an effective action \cite{string}.

In this article, we are investigating the warm inflationary scenario
in $F(T)$ theory. Previously in literature, generic inflation
scenarios have been investigated in DGP gravity model \cite{del
Campo:2007zj} and Brans-Dicke gravity theory \cite{Wu:2012yg}, to
name a few. It is correct that $F(T)$  gravity generates an
inflationary scenario but it is cold while we here study the warm
scenario which is far to our knowledge, has not been done before.
%%%%%%%%%%%%%%%%%%%
\section{The Model}
%%%%%%%%%%%%%%%%%%%%%
In literature there are a number of ways available to achieve the
inflation in the very early Universe including models involving
single field and multi-fields inflation, non-standard kinetic terms,
vector fields and nontrivial gravitational couplings. However we
study inflation via a single scalar inflaton in $F(T)$ gravity and
propose a Lagrangian as (units adopted for calculations are $16\pi
G=\hbar=c=1$)\footnote{Strictly speaking, $F(T)$ is not a model of
modified gravity like $F(R),F(R,G)$ and etc. It doesn't indicate any
simple modification of Einstein-Hilbert action. In a more common
form, it defines a geometry using asymmetric connections and based
on the idea of teleparallelism. Commonly it stated that the later
case is identical to GR. But it's only at the level of action.}
\begin{equation}\label{S'}
\mathcal{S}=\int
d^4x~e(\mathcal{L}_F+\mathcal{L}_\gamma+\mathcal{L}_\phi+\mathcal{L}_\text{int}),
\end{equation}
where $e=\text{det}(e^i_\mu)=\sqrt{-g}$, $e_i(x^\mu)$ are related to
the metric via $g_{\mu\nu}=\eta_{ij}e_\mu^i e^j_\nu$. where all
indices run over 0,1,2,3. $\mathcal{L}_F,$ $\mathcal{L}_\gamma$ and
$\mathcal{L}_\phi$ represent the Lagrangians for gravity model,
energy-matter and the inflaton scalar field  respectively. The last
term $\mathcal{L}_\text{int}$ plays the role  of the interaction
between inflaton as the scalar player of the inflation and other
fields. This scenario for inflation is called  warm inflation
\cite{Barrow,bary}. Specifically the total action reads
\begin{equation}\label{S}
\mathcal{S}=2\pi^2\int dt~ a(t)^3\Big[F(T)-\rho_\gamma+\frac{1}{2}
\phi_{,\mu}\phi^{,\mu}+U(\phi)+\mathcal{L}_\text{int}\Big].
\end{equation}
Here $a$ is the scale factor while $H\equiv\dot a/a$ is the Hubble
parameter. The inflaton $\phi$ has the potential energy $U(\phi)$
(to be determined in the later sections) and $\rho_\gamma$ is the
energy density of radiation component. The system of the field
equations is
\begin{eqnarray}
S^{\;\;\nu\rho}_{\mu}\partial_{\rho}TF_{TT}+\left[e^{-1}e^{i}_{
\mu}\partial_{\rho}\left(ee^{\;\;\alpha}_{i}S^{\;\;\nu\rho}_{\alpha}
\right)+T^{\alpha}_{\;\;\lambda\mu}S^{\;\;\nu\lambda}_{\alpha}\right]F_{T}
+\frac{1}{4}\delta^{\nu}_{\mu}F=\mathcal{T}^{\nu}_{\mu}\\
\nabla_{\mu}\nabla^{\mu}\phi+U'(\phi)=0.
\end{eqnarray}
The torsion scalar $T$ is defined by
\begin{equation}\nonumber
T=S^{\:\:\:\mu \nu}_{\rho} T_{\:\:\:\mu \nu}^{\rho}\,,
\end{equation}
and the components of the torsion tensor
$$
T_{\:\:\:\mu \nu}^{\rho}=e_i^{\rho}(\partial_{\mu}
e^i_{\nu}-\partial_{\nu} e^i_{\mu})\,,
$$
$$
S^{\:\:\:\mu \nu}_{\rho}=\frac{1}{2}(K^{\mu
\nu}_{\:\:\:\:\:\rho}+\delta^{\mu}_{\rho} T^{\theta
\nu}_{\:\:\:\theta}-\delta^{\nu}_{\rho} T^{\theta
\mu}_{\:\:\:\theta})\,,
$$
and also for contorsion tensor
$$
K^{\mu \nu}_{\:\:\:\:\:\rho}=-\frac{1}{2}(T^{\mu
\nu}_{\:\:\:\:\:\rho}-T^{\nu \mu}_{\:\:\:\:\:\rho}-T^{\:\:\:\mu
\nu}_{\rho})\,.
$$
Here $\mathcal{T}_{\mu\nu}=e^{a}_{\mu}\mathcal{T}_{a\nu}$ denotes
the energy-momentum tensor for matter field's Lagrangian $ L_m$, and
it is defined by
$$
\mathcal{T}_{a\nu}=\frac{1}{e}\frac{\delta L_m}{\delta e^{a\nu}}\,\,.
$$
Also in the KG equation, the covariant derivatives are with respect
to  the metricity condition $g^{\mu\nu}_{~~;\mu}=0$.

Note that the precise form of the interacting Lagrangian
$\mathcal{L}_\text{int}$ is not known, however a dissipation term
$\Gamma\dot\phi$ is introduced in the dynamical field equation of
the inflaton \cite{bary}. Earlier $\Gamma$ was considered as a
phenomenological function, but very recently starting from the first
principles, the general dissipation coefficient in low-temperature
warm inflation has been derived \cite{BasteroGil:2012cm}. The
implications of the above model were recently explored in \cite{ns}
via Noether symmetry approach and in \cite{ec} via energy
conditions. In \cite{ns} we showed that such a model admits
$F(T)\sim T^{3/4}$, $V(\phi)\sim \phi^2$ and can drive cosmic
acceleration in the late time evolution of the Universe while
crossing the phantom divide line at the present time. In the present
article, we discuss its implications in the very early Universe,
particularly cosmological inflation. The Friedmann-Robertson-Walker
(FRW) metric representing  a spatially flat, homogeneous and
isotropic spacetime is
\begin{equation}\label{frw}
ds^2=-dt^2+a(t)^2(dx^2+dy^2+dz^2).
\end{equation}
The modified Friedmann equations are \cite{ft1} (for a review see \cite{Bamba:2012cp})
\begin{eqnarray}
12H^2F_T+F=\rho_{\phi}+\rho_\gamma,\label{frw1}\\
48H^2\dot{H}F_{TT}-4F_T(3H^2+\dot{H})-F=p_{\phi}+p_\gamma,\label{frw2}
\end{eqnarray}
The torsion scalar reads as $T=-6H^2$, and associated inflaton's
density and pressure are
\begin{eqnarray}
\rho_{\phi}=\frac{1}{2}\dot{\phi}^2+U(\phi)\label{rho},\\
p_{\phi}=\frac{1}{2}\dot{\phi}^2-U(\phi)\label{p},
\end{eqnarray}
where $\rho_\phi$, $p_\phi$, show  the energy density and effective
pressure of inflaton, while $\rho_\gamma$ and $p_\gamma$ are
correspondingly energy density and pressure of the radiation. The
KG equation defines a unique causal  dynamics for the
scalar field with a frictional term \cite{bary}
\begin{equation}
\ddot{\phi}+3H\dot{\phi}+U'(\phi)=-\Gamma\dot{\phi}\label{kg},
\end{equation}
where we absorbed the interaction of inflaton with all other
existing fields during inflation using the dissipation factor
$\Gamma$. We mention here that from the dynamical point of the view,
Warm inflation is a phase transition with a dissipation auxiliary
sub-dominant mechanism. There is a possibility to compute this
factor for different fields (scalar, fermion) using quantum field
theory \cite{jcap2011}. Note that equation (\ref{kg}) is
inhomogeneous due to presence of a decay or dissipation term on the
right hand side. The term $-\Gamma\dot\phi$  in (\ref{kg}) shows the
decaying nature of the field and its conversion to the thermal
component. Eq.(\ref{kg}) is a special case of the generalized
Langevin equation \cite{ber1}. In general $\Gamma$ is a dynamical
parameter and can not be taken constant, but later we will
investigate a regime in which the form of $H(t)$ allows to take the
rate of dissipation as a very slowly varying function of time. It
turns out that cosmic inflation was de Sitter-like with
$\rho_\phi+p_\phi\approx0$ (or $w_\phi\approx-1$) \cite{mukh}. To
model a cosmic accelerated expansion in very early Universe, it is
convenient to use scalar fields and we employ the same strategy.
Such cosmic inflaton is quasi-stable and can decay to other forms of
energy like radiation and matter in the process of reheating as
discussed in conventional inflationary models. Moreover the quantum
fluctuations in a quantum inflaton will serve as seeds for later
structure formation. The dynamical equations are described by:
\begin{eqnarray}
\dot\rho_\phi+3H(\rho_\phi+p_\phi)&=&-\Gamma\dot\phi^2,\label{red1}\\
\dot\rho_\gamma+4H\rho_\gamma&=&\Gamma\dot\phi^2.\label{red2}
\end{eqnarray}
Here $\Gamma=f(\phi,T_A)$ may be taken general as a function of the
inflaton $\phi$, or the average temperature $T_A$, or both
\cite{ber}. Furthermore we suppose that $\Gamma>0$, as the Second
Law of Thermodynamics must be valid even in the inflationary era. In
inflationary epoch, the Friedmann equation (\ref{frw1}) reduces to
\begin{equation}\label{red}
12H^2F_T+F\approx\rho_{\phi}.
\end{equation}
The perturbations of the scalar field at the quantum level were the
initial seeds in the Universe and which culminated in the large
scale structure formation \cite{rocky}. It is believed that quantum
field theoretic version of warm inflation resolves the horizon and
flatness problem \cite{qft}. We consider $R$ as the decay (or
dissipation) rate defined by
\begin{equation}
R=\frac{\Gamma}{3H}.
\end{equation}
Here the decay rate $R$ is a dynamical quantity. To estimate $R$ we
need the form of $\Gamma$.  In accordance to the results of the QFT
the interacting supersymmetric theory \cite{sym}, the decay rate can
be written as
\begin{equation}
\Gamma\simeq C_\phi\frac{T_*^3}{\phi^2}.\label{gamma}
\end{equation}
where $C_\phi=0.64h^4\mathcal{N}$ in which
$\mathcal{N}=\mathcal{N}_\chi\mathcal{N}^2_{decay}$. Here
$\mathcal{N}_\chi$ is related to the superfield's decay (further
details can be seen in \cite{decay} and references therein). It  is
one possible form of the dissipation parameter. Other forms are
different and they need separate calculations (beyond the scope of
our work). We study two models:
\begin{eqnarray}
\Gamma\simeq\frac{T_*^3}{\phi^2} ,\ \ \text{Model-I}.\\
  \Gamma\simeq\Gamma_1=\text{const},\ \ \text{Model-II}.
\end{eqnarray}
Here $T_*$ is bath's temperature. Just as a historical fact we
mention here that in the original proposal of warm inflation, the
authors at that time, could not specify the precise form of $\Gamma$
due to the lack of a detailed model \cite{bary}. Later on taking
motivation from super-string theory and quantum field theoretic
approach, a suitable forms of $\Gamma$ were derived \cite{sym,
jcap2011,BasteroGil:2012cm}.\par
 From Eqs.
(\ref{red1}) and (\ref{red}), we obtain
\begin{eqnarray}
\dot{\phi}^2=-\frac{4\dot{H}(F_T+2TF_{TT})}{R+1}.\label{dotphi}
\end{eqnarray}
Here the choice of $F(T)$ must be astrophysically viable which means
that it must be in conformity with the observational data. We assume
that in inflationary epoch, radiation is prone to decay on average
on certain cosmic time scales i.e. $\dot\rho_\gamma\ll
4H\rho_\gamma$, and $\dot\rho_\gamma\ll \Gamma\dot\phi^2$. From
(\ref{red2}) we get the modified radiation density as
\begin{eqnarray}
\rho_\gamma=\frac{\Gamma \dot{\phi}^2}{4H}.\label{rhor}
\end{eqnarray}
In other words, if the decay is much rapid than cosmic expansion
$\Gamma\gg H$, (keeping $\dot\phi^2>0$ to avoid ghosts) the
radiation density will keep on increasing. In warm inflationary
scenario, the period of inflation continuously (but very rapidly)
dilutes to the radiation density. But due to presence of
dissipation, the depletion of radiation is compensated. Inserting
(\ref{dotphi}) in (\ref{rhor}) we obtain
\begin{eqnarray}
\rho_\gamma=-\frac{\Gamma\dot{H}(F_T+2TF_{TT})}{H(R+1)},\label{rhor2}
\end{eqnarray}
which can be written as $\rho_\gamma=C_{\gamma}T_*^4,$ where
$C_\gamma=\pi^2 g_*/30$. Here $g_*$ denotes the countable numbers of
relativistic degeneracy of states and  $T_*$ is the average
temperature for the background of thermal bath given by
\begin{equation}
T_*=\Big(-\frac{\Gamma\dot{H}(F_T+2TF_{TT})}{HC_\gamma(R+1)}\Big)^{1/4}.\label{T}
\end{equation}

Now from (\ref{frw1}) and (\ref{rhor2}) we obtain the inflaton
scalar potential
\begin{eqnarray}\label{U}
U(\phi)&=&-2TF_T+F+\frac{\dot{H}}{R+1}(F_T+2TF_{TT})\Big(2+\frac{\Gamma}{H}\Big).
\end{eqnarray}
We consider the following model of modified gravity
\cite{Jamil:2012nma}
\begin{eqnarray}
F(T)=T+\alpha\sqrt{-T}+\beta. \label{ft}
\end{eqnarray}
The first linear term indicates the Lagrangian for $F(T)$ gravity
and is dynamically equivalent to the general relativity at the level
of action. The second term denotes a ghost dark energy and performs
a role of cosmological constant and the last term $\beta$ is a
constant. We mention here that $F(T)$ represents the geometrical
part of our model. There is no dark energy in our model. Strictly
speaking, in the warm inflation scenario and in inflationary era
when the inflation is driven by inflaton field, we can safely
neglect the dark energy density. So, the only thing which we need is
a background with a definite geometry, here is the spacetime with
torsion and radiation and inflaton. So, this $F(T)$ can be
considered just as geometry. It is not necessary for $F(T)$ model to
mimic the equation of state of any kind of dark energy. But it can
be reconstructed mathematically as a toy model of a type of dark
energy ghost dark energy if and only if we neglect all matter
fields. In our context, we used this form of $F(T)$ because it is a
viable model. However we redefine the field equation in terms of
effective FRW field equations as the following:
\begin{eqnarray}
3H^2=\rho_{F(T)}+\Sigma\rho_{i},\ \ \rho_{F(T)}=-F(T)-T+2TF_{T}(T).
\end{eqnarray}
We observe that in the absence of any matter field (beyond our
inflationary scenario with matter fields like radiation,inflaton)
(\ref{ft}) gives us $\rho_{F(T)}=\beta$. It has the meaning of a
mathematical reconstruction of a type of dark energy which is not
interesting for us in this paper.  If we investigate dark energy
this form of energy density makes us worry. But here in inflationary
era, we do not  need any dark energy component. Further more
$\rho_{F(T)}$ does not indicate any dark energy. So absence of such
variable $\rho_{F(T)}$ does not sense any problem for our model. We
used $F(T)$ just for gravity of the model. For dynamical evolution
we have inflaton and radiation. Note that in inflationary
models,radiation and inflaton are the most important fields. The
action ansatz (\ref{ft}) is different form a pure cosmological
constant. It can be reconstructed from different points of view
\cite{H1,H2,H3,kk,reconstruction}. So this form of $F(T)$ is one of
the most physically viable models    of $F(T)$ proposed by us as the
first time \cite{Jamil:2012nma}. It reduces to the TEGR with a
cosmological constant. Even if we treat with $F(T)$ as a model of
dark energy this simple model is able to explain warm inflation in a
generalized teleparallel gravity.

In order to recover the intermediate scenario, we make  the scale
factor
\begin{eqnarray}
a(t)&=&A\exp\Big[ X(1+\omega(t-t_0))^Y \Big],\\
X&=&\frac{H_0(2n+1)}{2\omega(1+n)},\quad
Y=\frac{2n+2}{2n+1},\nonumber
\end{eqnarray}
which is very similar to the intermediate inflationary scenario
\cite{inter}. Since we do not need any further assumption on
acceleration in this inflationary era, we choose $t_0=0$. Thus our
ansatz  is viable as it yields an intermediate inflationary scenario
in $F(T)$ gravity. Using this scale factor we have:
\begin{eqnarray}
\dot{H}=\frac{c}{(-T)^n}
\end{eqnarray}
where $c<0$ and $n$ are constants. Indeed this equation is in agreement with $T=-6H^2$ if and only if $a(t)=A\exp\Big[ X(1+\omega(t-t_0))^Y \Big]$.,since:
\begin{eqnarray}
&&\dot{H}=\frac{c}{(-T)^n}\to \dot{H}H^{2n+1}=\frac{c}{6^n}\to H(t)=\Big(\frac{c(2n+1)}{6^n}(t-t_0)\Big)^{1/(2n+1)}\\&&\nonumber
\frac{d\log a(t)}{dt}=\Big(\frac{c(2n+1)}{6^n}(t-t_0)\Big)^{1/(2n+1)}\to a(t)=A\exp\Big[ X(1+\omega(t-t_0))^Y \Big],\\&&
X=\frac{H_0(2n+1)}{2\omega(1+n)},\quad
Y=\frac{2n+2}{2n+1},\nonumber.
\end{eqnarray}
 This is the same assumption which we used in our paper. To have a correct dimension, we
see  $\frac{c}{6^n}=\mathcal{O}(H_0^{2n+2})$, so it is adequate to
define a new parameter $k_1=\frac{c}{6^n H_0^{2n+2}}$, such that the
Hubble parameter reads as
\begin{eqnarray}
\dot{H}=k_1\frac{H_0^{2n+2}}{H^{2n}}, \ \ k_1<0\label{ansatz}.
\end{eqnarray}
The scale factor and Hubble parameter is suitably chosen so that  it
is consistent with the $F(T)$ gravity and the intermediate
expansion. The scale factor is necessary to perform the analysis and
therefore working with a hypothetical scale factor may not be
consistent with the inflationary scenario. Hence we picked the
intermediate scale factor which is also consistent with
astrophysical observations \cite{Barrow:2006dh}.

The slow-roll parameters which provide a necessary (but not
sufficient) condition for inflation are defined as
\cite{barrow,dany1,liddle}
\begin{equation}
\epsilon=-\frac{\dot H}{H^2},\quad
\eta=-\frac{\ddot H}{H\dot H},\label{slow}
\end{equation}
In the present context, using (\ref{ansatz}), the slow-roll
parameters yield
\begin{equation}
\epsilon=-k_1\Big(\frac{H_0}{H}\Big)^{2n+2},\quad \eta=2n
k_1\Big(\frac{H_0}{H}\Big)^{2n+2}.\label{epi}
\end{equation}
The first very interesting observation from (\ref{epi}) is that the
ratio between two slow-roll parameters remains constant, free from
the model of $F(T)$ or time evolution of the Hubble parameter i.e.
\begin{equation}
\frac{\eta}{\epsilon}=-2n.
\end{equation}
Also since always $H<H_0$ \footnote{The Hubble parameter after the
end of inflation is assumed to be larger compared to its value at
any later time.}, so to have a small set of the parameters we must
have $k_1<<1$. Since $\eta=\frac{m_\phi^2}{3H^2}$, our model implies
two possible situations: either $H\rightarrow\infty$ (it corresponds
to the $\epsilon\cong\eta \cong0$) and finite $m_\phi\neq0$ or
$m_\phi=0$ and $H$ is finite. Inflationary expansion stops when
$\rho_{\phi}\cong\rho_\gamma$ and the field violates the slow-roll
approximations $\dot H+H^2\approx 0,$ ($\ddot a\approx 0$) or
$\epsilon(\phi_{e})=1$.

In high dissipative regime  for our $F(T)$ model, we have:
\begin{eqnarray}
\rho_\gamma=-3\dot{H}(F_T+2TF_{TT}),\label{rhor2}
\end{eqnarray}

The condition of the inflation implies that $\ddot{a}<1$. It
corresponds to that $\epsilon<1$. Now we compute the number of
e-folding using the standard definition:
\begin{eqnarray}
N(t)=\int\limits_{t_1}^{t_2}H(t)dt.
\end{eqnarray}
where $t_1$ and $t_2$ correspond to time of start and end of
inflationary period respectively. We need also a perturbation theory
for our model. In the flat FRW background,  this perturbation can be
computed just using the perturbation of the inflaton $\phi$.  In the
warm inflation the expression of the fluctuations of the inflaton
reads
\begin{eqnarray}
<\delta \phi>_{\text{thermal}}=\Big(\Gamma H T_{*}^2\Big)^{1/4}.
\end{eqnarray}
Now for power spectrum of the primordial cosmic radiation, a more
general expression reads as the following \cite{BasteroGil:2009ec}
\begin{eqnarray}\label{pr}
P_R=R^{1/2}\Big(\frac{H^2}{2\pi \dot{\phi}}\Big)^2
\frac{T_*}{H}.
\end{eqnarray}
Here, the starred quantities correspond to the  evaluated parameters
at horizon crossing \cite{Bartrum:2013fia}. As mentioned in
\cite{Bartrum:2013fia}, that in the regime when particle production
is not so high, it is possible to ignore $n_*$ from the power
spectrum calculations. This power spectrum  is a function of the
wavelength. From observational data of WMAP7 \cite{WMAP7} we know
that for $k=0.002 Mpc^{-1}$ the following value is accepted:
\begin{eqnarray}
P_R=(2.445\pm0.096)\times 10^{-9}.
\end{eqnarray}

The scalar spectral (or power spectrum) index $n_s$ is given
by\footnote{Alternative definitions of $n_s$ are
$n_s=1+2\eta-6\epsilon$, and $n_s=1+4\frac{\dot H}{H^2}-\frac{\ddot
H}{H\dot H}$ \cite{guth1}.}
\begin{eqnarray}
n_s-1=\frac{d\ln P_R}{d\log k}.\label{ns-1}
\end{eqnarray}
The conventional inflationary models predict that the initial
density perturbations have a Gaussian distribution, and their power
spectrum index is, $n_s\approx1$. The early COBE experiment results
turned out to be in agreement with the above prediction
\cite{smoot}. As is well-known in the literature, the necessary
number of e-fold should be between $60-80$, to produce our
observable Universe \cite{guth}. For perfect Gaussianity ($n_s=1$),
however CMB is not completely Gaussian and the non-Gaussian (i.e.
$n_s$ deviates from unity by small amounts) features point to some
deeper physical mechanisms that still need to be understood
\cite{smith}. Our model predicts a non-Gaussianity of CMB which
varies for differently picked e-folding numbers. Here $d\ln
k(\phi)=dN(\phi)=(H/\dot\phi)d\phi$ i.e. wave number interval is
related with number of e-fold parameter. Following \cite{monti}, if
the tensor perturbations are generated during inflation, than it
would produce gravitational wave. The detection of these primordial
gravitational waves is already underway by LISA, BBO and DECIGO
\cite{sasaki} and also the value of Planck data $n_s\approx0.96$
\cite{planck}. The corresponding gravitational wave power spectrum
becomes
\begin{eqnarray}
P_g(\phi)=4\Big(\frac{H}{2\pi}\Big)^2.
\end{eqnarray}
Single inflaton slow roll scenario is falsifiable on the basis of the
following future observations: (1) CMB has large non-Gaussian
features, (2) Non-zero iso-curvature
perturbations and (3) Large running of the scalar spectrum. The
scalar-tensor ratio $r(\phi)$ is given by
\begin{eqnarray}
r(\phi)=\frac{P_g}{P_R}.\label{r1}
\end{eqnarray}
Also the spectral index for tensor perturbations is:
\begin{eqnarray}
n_t=-2\epsilon \label{nt}.
\end{eqnarray}
It is adequate here to  compare $n_s,$ and $r$ with the
observational data of pre-Planck \cite{planck}. As it is analyzed
through Planck, we know that \cite{planck,observationplanck}:
$n_s\cong 09603\pm0.0073$ (at $68\%$ confidence limit), $n_s\cong
0.961\pm0.007$ from pre-Planck data and finally $r<0.11$ (at $95\%$
confidence limit) and $r<0.22$ from WMAP7.

%%%%%%%%%%%%%%%%%%%%%%%%%%%%%%%%
\section{Intermediate Inflation}
%%%%%%%%%%%%%%%%%%%%%%%%%%%%%%%%%%%
As we mentioned before our $F(T)$ based model of inflation naturally
leads to the scale factor as the following Hubble parameter:
\begin{eqnarray}
H=H_0(1+\omega t)^{(2n+1)^{-1}},\ \ \omega=k_1 H_0(2n+1) \label{H(t)}.
\end{eqnarray}
This is similar to the intermediate form of  the Hubble parameter if
we shift the time and identify the parameters $H_0,$ $\omega$ with
the parameters $A,$ $f$ of the intermediate inflation model. For our
model, the e-folding number reads as
\begin{eqnarray}
N=\int_{t_1}^{t}{ H dt}=\frac{1}{2k_1 (1+n)}\Big((1+\omega t)^{(2n+2)/
(2n+1)}-(1+\omega t_1)^{(2n+2)/(2n+1)}\Big)\label{N}.
\end{eqnarray}
Here $t_1$ denotes the initializing time of the inflation and $t$ the ending time.

%%%%%%%%%%%%%%%%%%%%%%%%%%%%%%%%%%%%%%%%%%%%%%%%%%%%%%%%
\subsection{Model-I:  $\Gamma =\Gamma_0 \frac{T_*^3}{\phi^2}$}
%%%%%%%%%%%%%%%%%%%%%%%%%%%%%%%%%%%%%%%%%%%%%%%%%%%%%%
In slow roll approximation and in the high dissipation regime
$R>>1$, when (\ref{red}) is valid, using (\ref{gamma}) and (\ref{T})
and by using (\ref{H(t)}) we have the following solutions for $H$,
inflaton $\phi$ :
\begin{eqnarray}
\phi(t)=\phi_0 e^{\theta(1+\omega t)^{(4n+3)/(2(2n+1))}}.\\
H=H_0 \theta^{-2/(4n+3)}\Big(\log{\frac{\phi}{\phi_0}}\Big)^{2/(4n+3)}.
\end{eqnarray}
Here $\tau_0^{-2}=\frac{4 C_{\gamma} H_0 3^{1/4}}{\Gamma_0}$ ,
$\theta=\pm \frac{2(2n+1)}{\tau_0(3+4n)k_1 H_0}$ and
$\Gamma_0=\text{const}$. Using (\ref{epi}),(\ref{slow}) the slow
roll parameters reads are written as
\begin{eqnarray}
\epsilon=-k_1\theta ^{2(2n+2)/(4n+3)}\Big(\log{\frac{\phi}{\phi_0}}\Big)^{-2(2n+2)/(4n+3)}.\\
 \eta=2n k_1\theta ^{2(2n+2)/(4n+3)}\Big(\log{\frac{\phi}{\phi_0}}\Big)^{-2(2n+2)/(4n+3)}.
\end{eqnarray}

The energy density corresponds to the radiation is given by
\begin{eqnarray}
\rho_\gamma=\hat{\rho}\Big(\log{\frac{\phi}{\phi_0}}\Big)^{-3/(4n+3)}.
\end{eqnarray}
Where $\hat{\rho}=\frac{\Gamma_0 3^{3/4} H_0^{1/2}(-k_1/C_{\gamma})^{3/4}}{4\tau_0^2} \theta^{-3/(4n+3)}  $.\par
Number of e-folding reads from the (\ref{N}) as the following:
\begin{eqnarray}
N=\frac{\theta^{-2(2n+2)/(4n+3)}}{2k_1 (1+n)} \Big(\Big(\log{\frac{\phi}
{\phi_0}}\Big)^{2(2n+2)/(4n+3)}-\Big(\log{\frac{\phi_1}{\phi_0}}\Big)^{2(2n+2)/(4n+3)}\Big)\label{N}.
\end{eqnarray}
At the beginning of the inflation using $\epsilon(\phi_1)=1$ we
have:
\begin{eqnarray}
\phi_1=\phi_0 e^{\theta (-k_1)^{(4n+3)/(2(2n+2))}}.
\end{eqnarray}
Now we will compute the e-folding in terms of the inflaton (or vise vera):
\begin{eqnarray}
\phi=\phi_0 e^{(\frac{N}{N_c}-u)^{(4n+3)/(2(2n+2))}}
\end{eqnarray}
Here $N_c=\frac{\theta^{-2(2n+2)/(4n+3)}}{2k_1 (1+n)}$ and  $u=k_1 \theta^{\frac{2(2n+1)}{(3+4n)}}$.
\par
Now using this last equation, the expressions of the power-spectrum,
spectral index (scalar+tensor) and finally the tensor-scalar ratio
can be calculated using (\ref{pr},\ref{ns-1},\ref{r1},\ref{nt}) as
the following:
\begin{eqnarray}
P_R=P_0 (\frac{N}{N_c}-u)^{-\frac{1}{8}\frac{10n^2-15n-6}{(n+1)(2n+1)}} e^{-3(\frac{N}{N_c}-u)^{\frac{4n+3}{4(n+1)}}}.
\end{eqnarray}
\begin{eqnarray}
P_g=\frac{H_0^2 \theta^{-4/(4n+3)}}{\pi^2}(\frac{N}{N_c}-u)^{1/(n+1)}.
\end{eqnarray}
\begin{eqnarray}
n_s=1-\frac{H_0^{-\frac{1}{2}\frac{4n+1}{2n+1}}\theta^{\frac{4n+1}{(2n+1)(4n+3)}}}{\tau_0\phi_0}(\frac{N}{N_c}-u)^{-1/(2n+2)}e^{-(\frac{N}{N_c}-u)^{\frac{4n+3}{4(n+1)}}}\Big(\frac{1}{2}\frac{10n^2-15n-6}{(2n+1)(4n+3)}(\frac{N}{N_c}-u)^{-\frac{4n+3}{4(n+1)}}+3\Big)\label{ns(N)}.
\end{eqnarray}
and
\begin{eqnarray}
n_t=\frac{2k_1\theta ^{2(2n+2)/(4n+3)}}{\frac{N}{N_c}-u}.
\end{eqnarray}
Here
\begin{eqnarray}
P_0= \frac{\tau_0^2\Gamma_0^{1/2}}{4\pi^2\sqrt{3}\phi_0^3}\Big(-\frac{3k_1}{C_\gamma}\Big)^{5/8}H_0^{\frac{1}{4}\frac{30n+11}{2n+1}}\theta^{\frac{1}{2}\frac{10n^2-15n-6}{(2n+1)(4n+3)}}.
\end{eqnarray}
And also the scalar-tensor ratio reads:
\begin{eqnarray}
r(N)=\frac{H_0^2 \theta^{-4/(4n+3)}}{\pi^2 P_0} (\frac{N}{N_c}-u)^{\,{\frac {n+2+10\,{n}^{2}}{8 \left( n+1 \right)  \left( 2\,n+1
 \right) }}
}e^{3(\frac{N}{N_c}-u)^{(4n+3)/(2(2n+2))}}.\label{r(N)}.
\end{eqnarray}
Graphics of (\ref{ns(N)}),(\ref{r(N)}) are presented in figures 1
and 2. We set the parameters as $n=-\frac{3}{2}$ to compare with the
parameters in the usual intermediate inflationary models
$\omega=1,f=\frac{2n+2}{2n+1},A=\frac{(2n+1)H_0}{2n+2}$.
\begin{figure*}[thbp]
\begin{tabular}{rl}
\includegraphics[width=7.5cm]{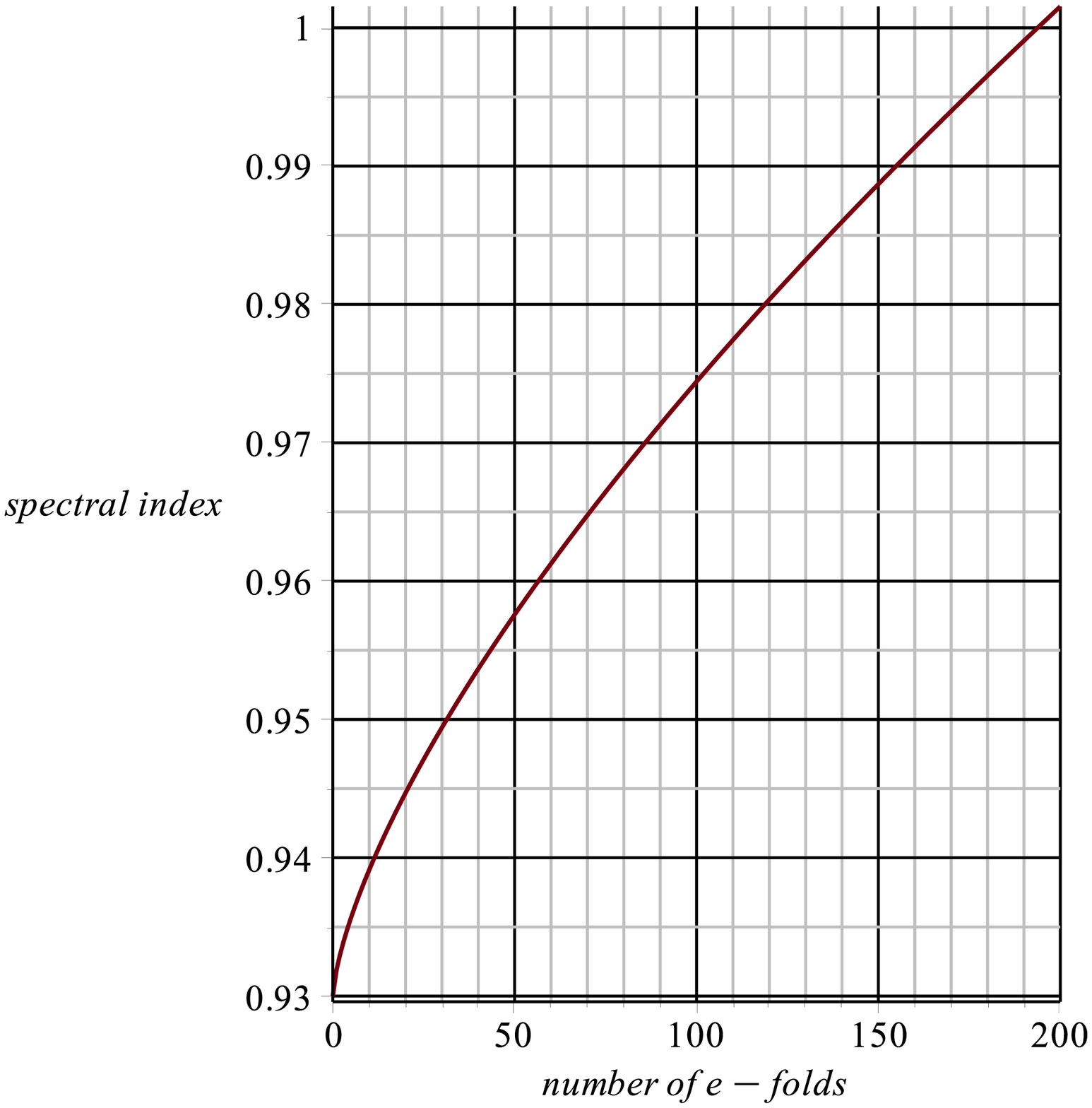}&
\includegraphics[width=7.5cm]{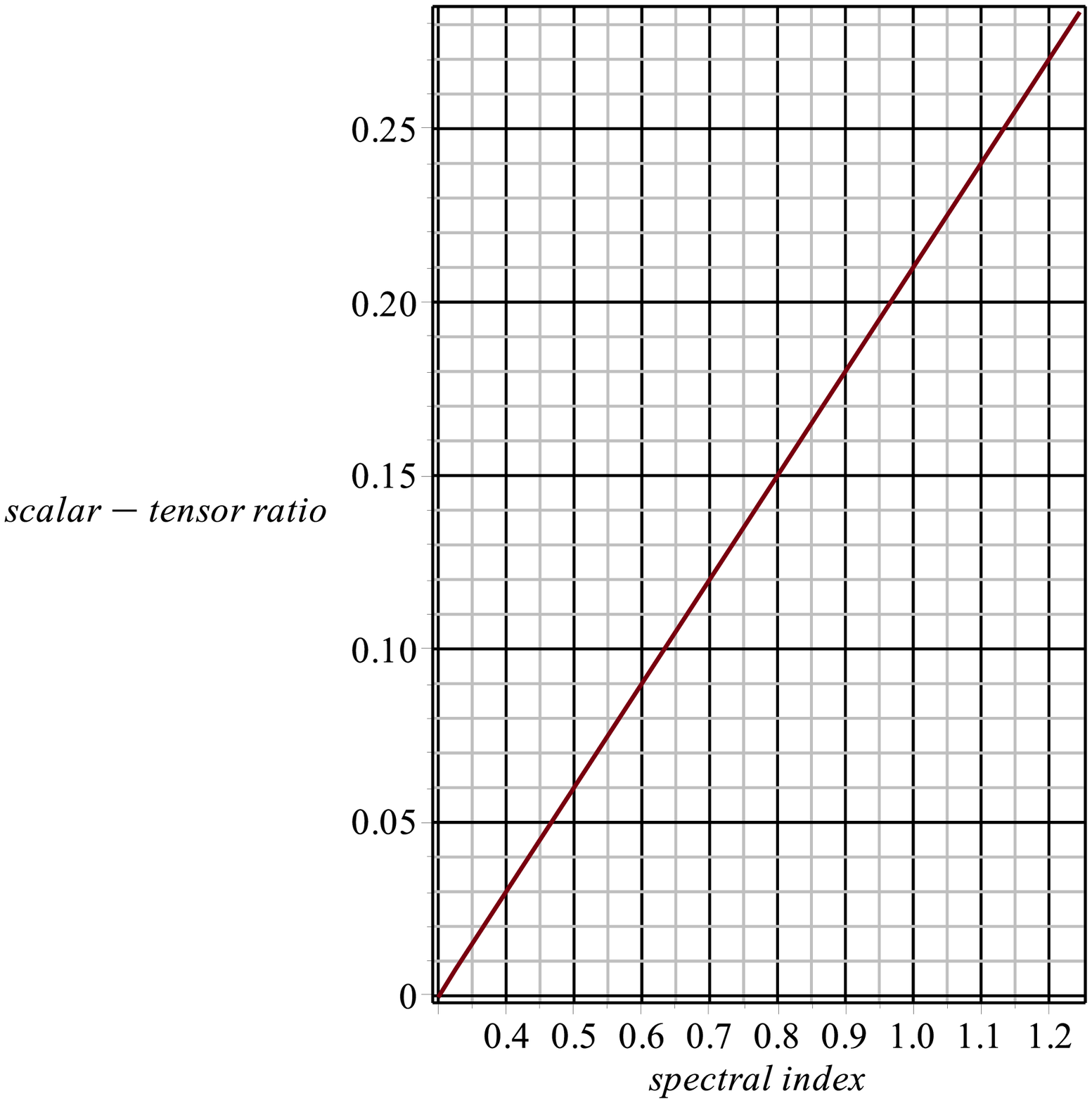} \\
\end{tabular}
\caption{  (\textit{Left}) Spectral scalar index $n_s$  versus N
given by (\ref{ns(N)}).   The data is confirmed by
$\Gamma_0=C_\gamma^{1/6}=70,n=-\frac{1}{5}, k_1\sim 7\times 10^7$.
Our model gives us $n_s(N=60)\sim0.95$ in a reasonable agreement
with the WMAP7 and for  pre Planck $n_s<0.961$. (\textit{Right})
Spectral index versus ratio of power spectra . We predict that
$n_s|_{0.69}\approx 0.12$ in agreement with observational data.}
\end{figure*}

%%%%%%%%%%%%%%%%%%%%%%%%%%%%%%%%%%%%%%%%%%%%%%%%%%%%%%%%
\subsection{Model-II:  $\Gamma \simeq \Gamma_1$}
%%%%%%%%%%%%%%%%%%%%%%%%%%%%%%%%%%%%%%%%%%%%%%%%%%%%%%
In this case and in the high dissipation regime , the solutions for inflaton and Hubble are written as the following:
\begin{eqnarray}
\phi=\phi_0+\theta_1(1+\omega t)^{\frac{1-2n}{2(1+2n)}}.\\
H=H_0\Big(\frac{\phi-\phi_0}{\theta_1}\Big)^{\frac{2}{1-n}}.
\end{eqnarray}
Here $\theta_1=\frac{2(2n+1)}{\tau_1\omega (2n+3)},\tau_1=\sqrt{-\frac{\Gamma_1(2n+1)}{12 H_0^2\omega}}$.
So, the slow role parameters read:
\begin{eqnarray}
\epsilon=-k_1\Big(\frac{\phi-\phi_0}{\theta_1}\Big)^{\frac{4(n+1)}{2n-1}}.\\
\eta=2nk_1\Big(\frac{\phi-\phi_0}{\theta_1}\Big)^{\frac{4(n+1)}{2n-1}}.
\end{eqnarray}
Also, the radiation density reads:
\begin{eqnarray}
\rho_\gamma=\frac{-3 H_0
\omega}{2n+1}\Big(\frac{\phi-\phi_0}{\theta_1}\Big)^{\frac{4n}{2n-1}}.
\end{eqnarray}
So, the e-folding from (\ref{N}) reads:
\begin{eqnarray}
N=\frac{1}{2k_1(n+1)}\Big[\Big(\frac{\phi-\phi_0}{\theta_1}\Big)^{\frac{4(n+1)}{1-2n}}-\Big(\frac{\phi_1-\phi_0}{\theta_1}\Big)^{\frac{4(n+1)}{1-2n}}\Big].
\end{eqnarray}
The starting inflaton's magnitude is :
\begin{eqnarray}
\phi_1=\phi_0+\theta_1(2nk_1)^{\frac{-(2n-1)}{4(n+1)}}.
\end{eqnarray}
So, it is possible to rewrite the inflaton in terms of the N,:
\begin{eqnarray}
\phi=\phi_0+\theta_1\Big[(2k_1(1+n)) N+\frac{1}{2nk_1}\Big]^{\frac{1-2n}{4(1+n)}}.
\end{eqnarray}
So, the temperature is obtained:
\begin{eqnarray}
T_*=\Big(\frac{-3 H_0 \omega}{C_\gamma}\Big)^{{1}{4}}\Big(\frac{\phi-\phi_0}{\theta_1}\Big)^{\frac{n}{2n-1}}
\end{eqnarray}

Now we can write the spectrum and indexes functions as the following list:
\begin{eqnarray}
P_R=P_1\Big[(2k_1(1+n)) N+\frac{1}{2nk_1}\Big]^{\frac{3}{4}}.
\end{eqnarray}
Here
\begin{eqnarray}
P_1=\frac{H_0^{5/2} \tau_1^2}{4\pi^2\sqrt{3}} \Gamma_1^{1/2} \Big(\frac{-3 H_0 \omega}{C_\gamma}\Big)^{\frac{1}{4}}.
\end{eqnarray}
\begin{eqnarray}
P_g=\frac{H_0^2}{\pi^2}\Big[(2k_1(1+n)) N+\frac{1}{2nk_1}\Big]^{\frac{1}{1+n}}.
\end{eqnarray}

\begin{eqnarray}
n_t=-4nk_1\Big[(2k_1(1+n)) N+\frac{1}{2nk_1}\Big]^{-1}.
\end{eqnarray}

The scalar to tensor ratio
\begin{eqnarray}
r=\frac{H_0^2}{\pi^2 P_1}\Big[(2k_1(1+n)) N+\frac{1}{2nk_1}\Big]^{-\frac{1+3n}{4(1+n)}}\label{r2}..
\end{eqnarray}
So, the index $n_s$ reads:
\begin{eqnarray}\label{ns2}
n_s=1-\frac{3(n+1)}{\theta_1\tau_1 H_0(2n-1)}\Big[(2k_1(1+n)) N+\frac{1}{2nk_1}\Big]^{\frac{-1}{2(1+n)}}.
\end{eqnarray}
We check the data for (\ref{r2},\ref{ns2}). We set $\Gamma_1=C_{\gamma}^{1/6},n=\frac{-5}{4}$.

\begin{figure*}[thbp]
\begin{tabular}{rl}
\includegraphics[width=7.5cm]{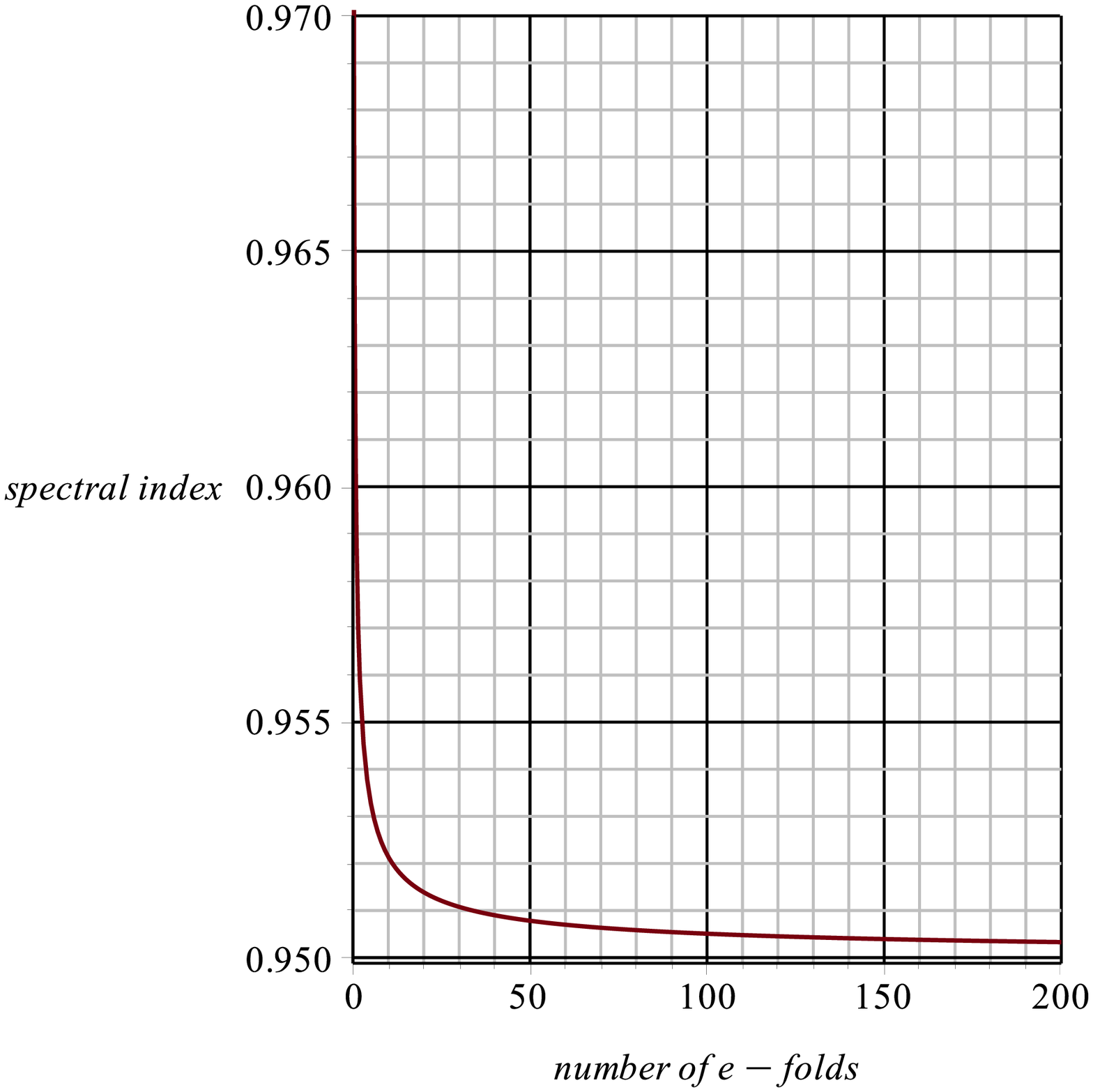}&
\includegraphics[width=7.5cm]{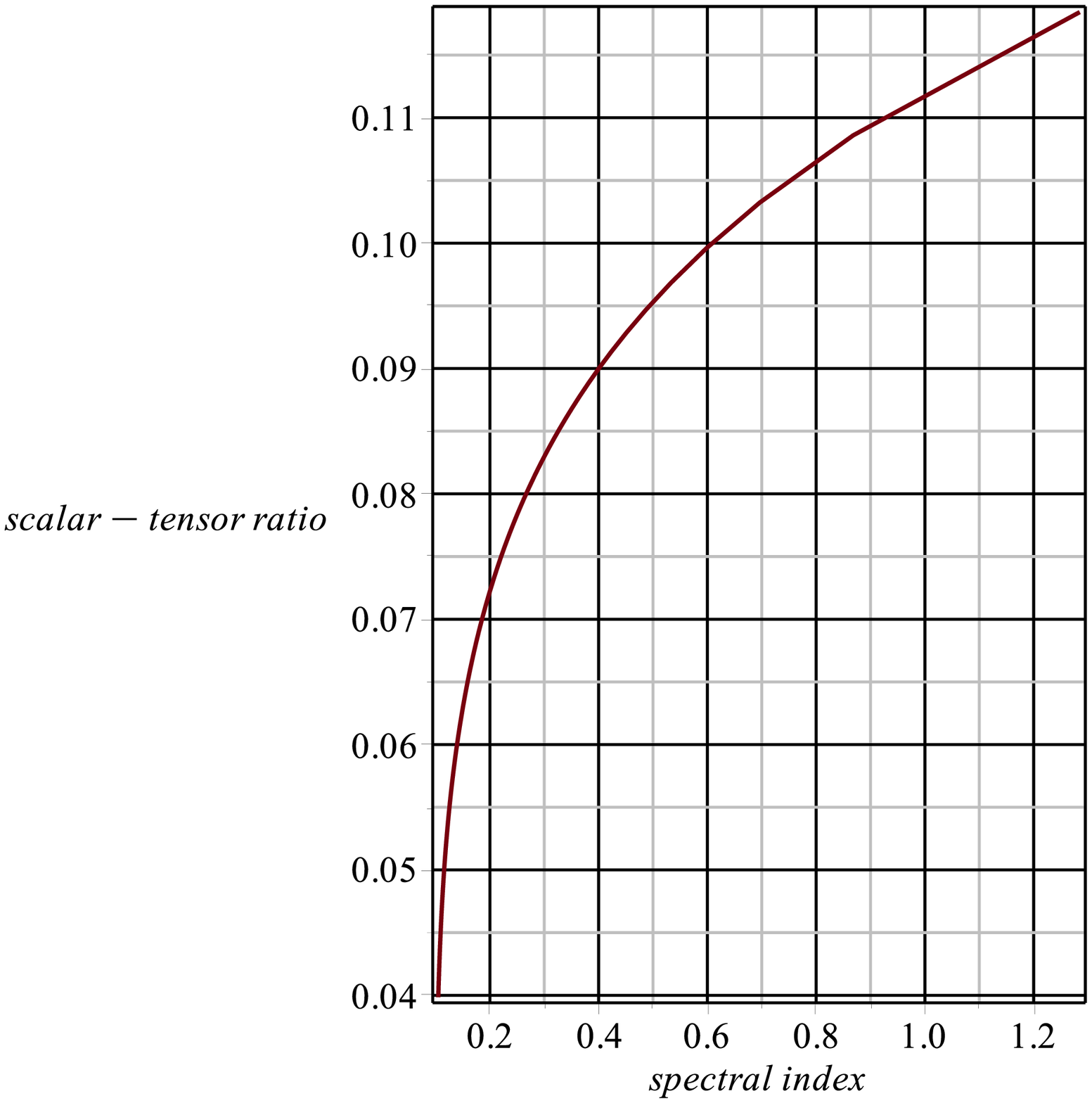} \\
\end{tabular}
\caption{ (\textit{Left}) Spectral scalar index $n_s$  versus N
given by (\ref{ns-1}).   The data is confirmed by
$\Gamma_0=C_\gamma^{1/6},C_\gamma^{1/6}=70,n=-\frac{1}{5}, k_1\sim
7\times 10^7$. Our model gives us $n_s(N=50)\sim0.95$ in a
reasonable agreement with the WMAP7 and for  pre Planck $n_s<0.97$.
(\textit{Right}) Spectral index versus ratio of power spectra . We
predict that $n_s|_{0.69}\approx 0.95$ in agreement with
observational data. }
\end{figure*}

\section{Conclusion}
In summary, for the first time in literature,  we investigated
cosmological warm inflation model in the framework of
teleparallel gravity. We introduced a canonical inflaton to act as
an inflaton field. We calculated the relevant inflationary
parameters such as scalar-tensor ratio, power-spectrum indices for
density perturbations and gravitational waves and e-folding
parameter. We get a constant ratio of two slow-roll parameters whose
value depends on our ansatz. Like generic inflationary models, ours
also predicts a gravitational wave background with a power spectrum.
Our calculations support the warm-intermediate inflationary
scenario. Moreover our results are compatible with compatible with
astrophysical observations of cosmic microwave background and pre
Planck. Our principal motivation is firstly we proposed warm inflation in torsion based spacetime,for the first time using a scalar field. But the main result which it differs our work from any previous work is that our numerical results and out proposed mechanism gives us $n_s(N=60)\sim0.95$ in a reasonable agreement
with the WMAP7 and for pre Planck $n_s<0.961$. So, in agreement with data we success to warm inflation in F(T). This is the main credit of the present work for publication. We mention here that a better and viable model of warm inflation can be investigate if we use a dynamical form of $F(T)=T+\alpha\sqrt{-T} +\sqrt{-T} Log{(-T)}+\beta$. It means we want $\rho_{F(T)}=\rho_{DE}$ to be proportional to $H(t)$. Although as we explained it in details, in our warm inflation we can safely neglect the effect of dark energy in favor of other dominant matter fields, but this suggestion will be carried out as a
separate paper.

\section{ACKNOWLEDGEMENTS}
We are indebted to  J. Barrow, M. Bastero-Gil, A.
Berera, R. O. Ramos, J. G. Rosa for  pointing out corrections and making suggestions for improvement.


\begin{thebibliography}{99}
\bibitem{guth} A. Guth, Phys. Rept. 333, 555-574 (2000); A. Guth, J. Phys. A 40, 6811
(2007); A. Linde, Lect. Notes Phys. 738, 1-54 (2008).
\bibitem{cole} A. Lewis, A. Challinor, Phys. Rept. 429, 1-65 (2006).
\bibitem{buchdahl} H. A. Buchdahl, Mon. Not. Roy. Astron. Soc. 150, 1 (1970).
\bibitem{lorenzo}
 O.~Gorbunova and L.~Sebastiani,
  %``Viscous Fluids and Gauss-Bonnet Modified Gravity,''
  Gen.\ Rel.\ Grav.\  42, 2873 (2010)
  [arXiv:1004.1505 [gr-qc]].
  %%CITATION = ARXIV:1004.1505;%%
  %8 citations counted in INSPIRE as of 10 Sep 2013
     D.~Momeni, N.~Majd and R.~Myrzakulov,
  %``p-wave holographic superconductors with Weyl corrections,''
  Europhys.\ Lett.\  {\bf 97}, 61001 (2012)
  [arXiv:1204.1246 [hep-th]].
  %%CITATION = ARXIV:1204.1246;%%
  %15 citations counted in INSPIRE as of 10 Sep 2013.
    %\cite{Jamil:2012zm}
  M.~Jamil, S.~Ali, D.~Momeni and R.~Myrzakulov,
  %``Bianchi Type I Cosmology in Generalized Saez-Ballester Theory via Noether Gauge Symmetry,''
  Eur.\ Phys.\ J.\ C {\bf 72}, 1998 (2012)
  [arXiv:1201.0895 [physics.gen-ph]].




\bibitem{review} S. Nojiri, S. D. Odintsov, Phys. Rept. 505, 59-144 (2011);  S.~'i.~Nojiri and S.~D.~Odintsov,
  %``Introduction to modified gravity and gravitational alternative for dark energy,''
  eConf C 0602061, 06 (2006)
  [Int.\ J.\ Geom.\ Meth.\ Mod.\ Phys.\   4, 115 (2007)]
  [hep-th/0601213].





\bibitem{hayashi} K. Hayashi, T. Shirafuji, Phys. Rev. D 19, 3524 (1979);
 K. Hayashi, T. Shirafuji, Phys. Rev. D 24, 3312 (1981).
\bibitem{ft1}
R. Ferraro , F.  Fiorini, Phys. Rev. D,
75, 084031 (2007), [arXiv:gr-qc/0610067];
G.R. Bengochea ,R. Ferraro , Phys. Rev. D, 79, 124019
(2009), [arXiv:0812.1205];
R. Ferraro ,F. Fiorini, Phys. Rev. D, 78,
124019 (2008), [arXiv:0812.1981];
E.V. Linder , Phys. Rev. D, 81,
127301 (2010), [arXiv:1005.3039].
\bibitem{ft11}
 R.~Myrzakulov,
  %``Accelerating universe from F(T) gravity,''
  Eur.\ Phys.\ J.\ C  71, 1752 (2011)
  [arXiv:1006.1120 [gr-qc]].
  %%CITATION = ARXIV:1006.1120;%%;
K. Bamba, R. Myrzakulov, S. Nojiri, S. D. Odintsov, Phys. Rev. D 85, 104036
(2012); L. Iorio, E.N. Saridakis, MNRAS 427, 1555 (2012)
\bibitem{ft2} M. Jamil, D. Momeni, R. Myrzakulov, Eur. Phys. J. C 72, 2122
(2012); M. Jamil, D. Momeni, R. Myrzakulov, Eur. Phys. J. C  72,
2075 (2012); M. Sharif, S. Azeem,  Astrophys. Space Sci.
342(2012)521-530.

\bibitem{paper} S. Basilakos, S. Capozziello, M. De Laurentis, A. Paliathanasis and M. Tsamparlis, arXiv:1311.2173v1 [gr-qc].

\bibitem{ft} M. E. Rodrigues, M. J. S. Houndjo, J. Tossa, D. Momeni, R. Myrzakulov, JCAP11(2013)024

\bibitem{Barrow}
J. D. Barrow,Phys. Lett. B 235, 40 (1990); J. D. Barrow, A. R. Liddle, C. Pahud,    Phys.Rev.D74:127305 (2006),arXiv:astro-ph/0610807.
\bibitem{qft}
M.Bastero-Gil,A.Berera,R.O.Ramos,J.Cosmol.Astropart.Phys.1109(2011)033,arXiv:1008.1929[hep-ph];
M.Bastero-Gil,A.Berera,R.O.Ramos,J.G.Rosa,J.Cosmol.Astropart.Phys.1301(2013)016,arXiv:1207.0445[hep-ph].
\bibitem{barrow1} J. D. Barrow, A. R. Liddle, Phys. Rev. D 47, 5219
(1993).
\bibitem{campo1} J. D. Barrow, Class. Quant. Grav. 13, 2965 (1996);
 A. Cid, S. del Campo, AIP Conf. Proc. 1471, 114 (2012).
\bibitem{string}
A.K.Sanyal,Phys.Lett.B645(2007)1,arXiv:astro-ph/0608104;T.Koivisto,D.F.Mota,Phys.Lett.B644(2007)104,arXiv:astro-ph/0606078;
T.Koivisto,D.F.Mota,Phys.Rev.D75(2007)023518,arXiv:hep-th/0609155.
\bibitem{del Campo:2007zj}
  S.~del Campo and R.~Herrera,
  %``Warm inflation in the DGP brane-world model,''
  Phys.\ Lett.\ B {\bf 653}, 122 (2007)
  [arXiv:0708.1460 [gr-qc]]
\bibitem{Wu:2012yg}
  Y.~-H.~Wu and C.~-H.~Wang,
  %``Brans-Dicke theory of gravity with torsion: A possible solution of $\omega$-problem,''
  Phys.\ Rev.\ D {\bf 86}, 123519 (2012)
  [arXiv:1210.1010 [gr-qc]].
 \bibitem{setare}
 M.~R.~Setare and V.~Kamali,
  %``Tachyon Warm-Intermediate Inflationary Universe Model in High Dissipative Regime,''
  JCAP 1208, 034 (2012)
  [arXiv:1210.0742 [hep-th]];
     M.~R.~Setare and V.~Kamali,
  %``Cosmological perturbations in warm-tachyon inflationary universe model with viscous pressure on the brane,''
  JHEP 1303, 066 (2013)
  [arXiv:1302.0493 [hep-th]];
    M.~R.~Setare and V.~Kamali,
  %``Tachyon Warm-Logamediate Inflationary Universe Model in High Dissipative Regime,''
  Phys.\ Rev.\ D  87, 083524 (2013)
  [arXiv:1305.0740 [hep-th]];
\bibitem{singleton}
S.~K.~Modak and D.~Singleton,
  %``Hawking Radiation as a Mechanism for Inflation,''
  Int.\ J.\ Mod.\ Phys.\ D  21, 1242020 (2012)
  [arXiv:1205.3404 [gr-qc]];  S.~K.~Modak and D.~Singleton,
  %``Inflation with a graceful exit and entrance driven by Hawking radiation,''
  Phys.\ Rev.\ D  86, 123515 (2012)
  [arXiv:1207.0230 [gr-qc]].

\bibitem{jamil} H. M. Sadjadi, M. Jamil, Gen. Rel. Grav. 43, 1759
(2011).
\bibitem{bary} A. Berera, L-Z Fang, Phys. Rev. Lett. 74, 1912 (1995).
\bibitem{BasteroGil:2012cm}
  M.~Bastero-Gil, A.~Berera, R.~O.~Ramos and J.~G.~Rosa,
  %``General dissipation coefficient in low-temperature warm inflation,''
  JCAP 1301, 016 (2013)
  [arXiv:1207.0445 [hep-ph]].

\bibitem{ns} M. Jamil, D. Momeni, R. Myrzakulov, Eur. Phys. J. C 72, 2137 (2012).
\bibitem{ec} M. Jamil, D. Momeni, R. Myrzakulov, Gen. Relativ. Grav. 45, 263
(2013).
\bibitem{jcap2011}
 M. Bastero-Gil, A. Berera, R. O. Ramos
JCAP 1109 (2011) 033,  arXiv:1008.1929 [hep-ph].



\bibitem{Bamba:2012cp}
  K.~Bamba, S.~Capozziello, S.~'i.~Nojiri and S.~D.~Odintsov,
  %``Dark energy cosmology: the equivalent description via different theoretical models and cosmography tests,''
  Astrophys.\ Space Sci.\  342, 155 (2012)
  [arXiv:1205.3421 [gr-qc]].
\bibitem{ber1}  A. Berera,   arXiv:hep-ph/9807523
\bibitem{mukh} V. Mukhanov, Physical Foundations of Cosmology,
Cambridge University Press (2005).
\bibitem{ber} A. Berera, Phys. Rev. Lett. 75, 3218 (1995); A. Berera, Phys. Rev. D 55, 3346 (1997).
\bibitem{rocky} E.W. Kolb, M.S. Turner, The Early
Universe (Addison-Wesley, New York, 1990).
\bibitem{qft} A. Berera, M. Gleiser, R. O. Ramos, Phys. Rev. Lett. 83, 264 (1999).

\bibitem{sym} I. G. Moss, C. Xiong, arXiv:hep-ph/0603266.
\bibitem{decay} J. C. B. Sanchez, M. Bastero-Gil, A. Berera, K. Dimopoulos, Phys. Rev. D 77, 123527
(2008).
\bibitem{Jamil:2012nma}
  M.~Jamil, D.~Momeni and R.~Myrzakulov,
  %``Attractor Solutions in $f(T)$ Cosmology,''
  Eur.\ Phys.\ J.\ C  72, 1959 (2012)
  [arXiv:1202.4926 [physics.gen-ph]].
  %%CITATION = ARXIV:1202.4926;%%
  %25 citations counted in INSPIRE as of 10 Sep 2013
\bibitem{H1}
F.R. Urban, A.R. Zhitnitsky, Phys. Rev. D 80, 063001 (2009).
\bibitem{H2} F.R. Urban, A.R. Zhitnitsky, Phys. Lett. B 688, 9 (2010).
\bibitem{H3}N. Ohta, Phys. Lett. B 695, 41 (2011).
\bibitem{kk}
K. Karami, A. Abdolmaleki, S. Asadzadeh, Z. Safari,Eur. Phys. J. C 73 2565,(2013).
\bibitem{reconstruction}
P.~Huang, Y.~-C.~Huang and F.~-F.~Yuan,
  %``$f(T)$ gravity from holographic Ricci dark energy model with new boundary conditions,''
  Mod.\ Phys.\ Lett.\ A {\bf 28}, no. 39, 1350171 (2013)
  [arXiv:1312.6762 [gr-qc]].

\bibitem{Barrow:2006dh}
  J.~DBarrow, A.~RLiddle and C.~Pahud,
  %``Intermediate inflation in light of the three-year WMAP observations,''
  Phys.\ Rev.\ D {\bf 74}, 127305 (2006)
  [astro-ph/0610807].[arXiv:astro-ph/0610807].
\bibitem{inter} A. D. Rendall, Class. Quant. Grav. 22, 1655 (2005); S. del Campo, R.
Herrera, Phys. Lett. B 670, 266 (2009);  R. Herrera, N. Videla, Eur.
Phys. J. C 67, 499 (2010).

\bibitem{barrow} J. D. Barrow, N. J. Nunes, Phys. Rev. D 76, 043501
(2007).
\bibitem{dany1} D. Baumann, D. Green, JHEP 1104, 071 (2011).
\bibitem{liddle} A. Liddle, D. Lyth, Cosmological Inflation and Large-Scale Structure (Cambridge
University Press, 2000); J. Linsey, A. Liddle, E. Kolb, E. Copeland,
Rev. Mod. Phys 69, 373 (1997); B. Bassett, S. Tsujikawa, D. Wands,
Rev. Mod. Phys. 78, 537 (2005); S.K. Modak, D. Singelton, Phys. Rev.
D 86, (2012) 123515; Int. J. Mod. Phys. D 21 (2012) 1242020.
\bibitem{BasteroGil:2009ec}
  M.~Bastero-Gil and A.~Berera,
  %``Warm inflation model building,''
  Int.\ J.\ Mod.\ Phys.\ A {\bf 24}, 2207 (2009)
  [arXiv:0902.0521 [hep-ph]].
\bibitem{Bartrum:2013fia}
  S.~Bartrum, M.~Bastero-Gil, A.~Berera, R.~Cerezo, R.~O.~Ramos and J.~G.~Rosa,
  %``The importance of being warm (during inflation),''
  arXiv:1307.5868 [hep-ph].
\bibitem{WMAP7}
E.Komatsu,etal.,arXiv:1001.4538[astro-ph.CO];B.Gold,etal.,arXiv:1001.4555[astro-ph.GA];
D.Larson,etal.,arXiv:1001.4635[astro-ph.CO].


\bibitem{guth1} A. Guth, astro-ph/0404546.
\bibitem{smoot} G. Smoot et al., Astrophys. J. 396, L1 (1992).
\bibitem{smith} G. Rossmanith, H. Modest, C. Raeth, A. J. Banday, K. M. Gorski, G. Morfill,
Advances in Astronomy, Volume 2011, 174873 (2011).
\bibitem{monti} K. Bhattacharya, S. Mohanty, A. Nautiyal, Phys. Rev. Lett. 97,
251301 (2006).
\bibitem{sasaki} L. Alabidi, K. Kohri, M. Sasaki, Y. Sendouda, JCAP 09, 017
(2012).
\bibitem{planck}
P. A. R. Ade et al. [Planck Collaboration], arXiv:1303.5076
[astro-ph.CO]; P. A. R. Ade et al. [Planck Collaboration],
arXiv:1303.5082 [astro-ph.CO].
\bibitem{damy} D. A. Easson, B. A. Powell, Phys. Rev. Lett. 106, 191302
(2011).

\bibitem{campo} S. del Campo, R. Herrera, JCAP 0904, 005 (2009); R. Herrera, E. San
Martin, Eur. Phys. J. C 71, 1701 (2011)
\bibitem{campo3} R. Herrera, M. Olivares,  arXiv:1205.2365v1
[gr-qc].



\bibitem{observationplanck}
C. Cheng, Q. -G. Huang and Y. -Z. Ma, JCAP 1307, 018
(2013) [arXiv:1303.4497 [astro-ph.CO]].
\end{thebibliography}
\end{document}